# TRANSPORT PROPERTIES OF MgB$_2$

M.Putti, E.Galleani d'Agliano, D.Marrè, F.Napoli, M.Tassisto,
INFM/CNR, Dipartimento di Fisica, Via Dodecaneso 33, 16146 Genova, Italy

P.Manfrinetti, A.Palenzona
INFM, Dipartimento di Chimica e Chimica Industriale, Via Dodecaneso 31, 16146 Genova, Italy

## 1. Introduction

The discovery of the superconductivity in MgB$_2$ [1] at a temperature as high as 39 K has stimulated a enormous interest about the comprehension of its mechanism [2][3].

The study of the electron transport properties may give an insight into the normal state conduction process and into the electronic structure of MgB$_2$. From this point of view the results are contradictory. In fact, some measurements, such as magnetoresistivity [4] and the Seebeck effect [5][6][7][8], indicate that MgB$_2$ behaves like a simple metal, while the Hall effect measurements [9] are very similar to those of the high temperature superconductors and the thermal conductivity [8][10] does not show any kind of anomaly at the superconducting transition. However, it should be stressed that most measurements have been performed on sintered samples in which grain boundaries can affect transport properties complicating a clear understanding of MgB$_2$ basic properties. Actually, since the coherence length, $\xi \approx 50$ Å, is larger than the junction length between the grains, the transport properties that involve the motion of Cooper pairs are less affected than in cuprates superconductors. Thus, the resistivity transitions are sharp and the absence of weak links has been emphasized by magnetization measurements [11][12]. On the other hand, when quasi-particles are responsible for the transport, their scattering mechanisms at the grain boundaries cannot be neglected and this could cause incorrect estimate of the intrinsic properties of $MgB_2$.

In this paper we present the resistivity, the Seebeck effect, and the thermal conductivity measurements on a $MgB_2$ sintered sample. Such transport properties highlight the role of the junctions between the grains to a different extent. In particular, the temperature dependence of resistivity may be explained with the assumption that grain boundaries have a resistance, independent of temperature, in series with the resistance of grains. Also the behaviour of the Seebeck effect, as long as the scattering at the grain boundaries is elastic, is not affected at all by granularity [13]. Thus, the thermopower (TEP) can be a useful tool to provide information on the electronic structure. On the other hand, the grain boundary resistance affects the thermal conductivity strongly, masking the superconducting transition completely. However, this complete set of transport properties gives us the possibility to separate the contribution of grains and several conclusions can be drawn. To summarize, we will show that $MgB_2$ behaves like a simple metal in which electron correlation does not play any role and the high phonon frequencies and the electron-phonon coupling are the main features that allow to account for all the transport properties very well.

## 2. Experimental

The compound $MgB_2$ was prepared by direct synthesis from the elements: Mg (in form of fine turnings, 99.999 wt.% purity) and crystalline B (325-mesh powder, 99.7 wt.% purity), were mixed together accurately and closed by arc welding under pure argon into outgassed Ta crucibles which were then sealed in quartz ampoules under vacuum. The samples were slowly heated up to 950 °C and maintained at this temperature for 1 day. The final product was a gray-blackish powder, only slightly sintered; it was characterized by *X*-ray diffraction using a Guinier camera and CuKα radiation (pure Si as an internal standard, *a*= 5.4308 Å). The value of the lattice parameters, *a*= 3.087(1), *c*= 3.526(1) Å, in very good agreement with the literature, and the absence of extra reflections in the diffraction pattern indicate a very pure product.

Specimens for physical measurements, in the shape of parallelepiped bar (2x3x12 $mm^3$), were prepared by pressing the powders in a stainless-steel die into a pellet which was then sintered by heat treatment at 1000 °C for 2 days (again in Ta containers, welded under argon and closed in silica tubes under vacuum). The density of the bars was about the 70% of the calculated value on the base of the lattice parameters. SEM images of the sintered sample (see

fig.1) show a rather homogeneous distribution of crystallites with an average dimension of the order of µm (fig. 1a); its magnification (fig.1b) shows the shape of grains which are platelets with a section in the *ab*-plane of few µm$^2$ and a thickness of the order of one tenth of µm in the *c*-direction.

The sintered sample used for transport measurements was characterized by a Quantum Design SQUID magnetometer and showed the transition onset at T=38.8 K.

The resistivity measurement were performed using a standard four-probe technique. The Seebeck effect was measured using an a.c. technique described elsewhere [14].The gradient applied to the sample was varied from 3 to 1 K/cm; these high values were necessary owing to the small values of the Seebeck effect. The frequency was chosen as low as ν=0.005-0.003 Hz in order to avoid a reduction of the heat wave amplitude along the sample. Under these conditions the thermal conductivity, defined as $\kappa = J(\nu)/\nabla T(\nu)$ (where $J(\nu)$ is the heat flow provided at the frequency ν and $\nabla T(\nu)$ is the temperature gradient oscillating at the frequency ν), was measured simultaneously with the Seebeck effect. For the measurements of *S* and $\kappa$ we estimate a sensitivity is of 0.5% and accuracy of 2%.

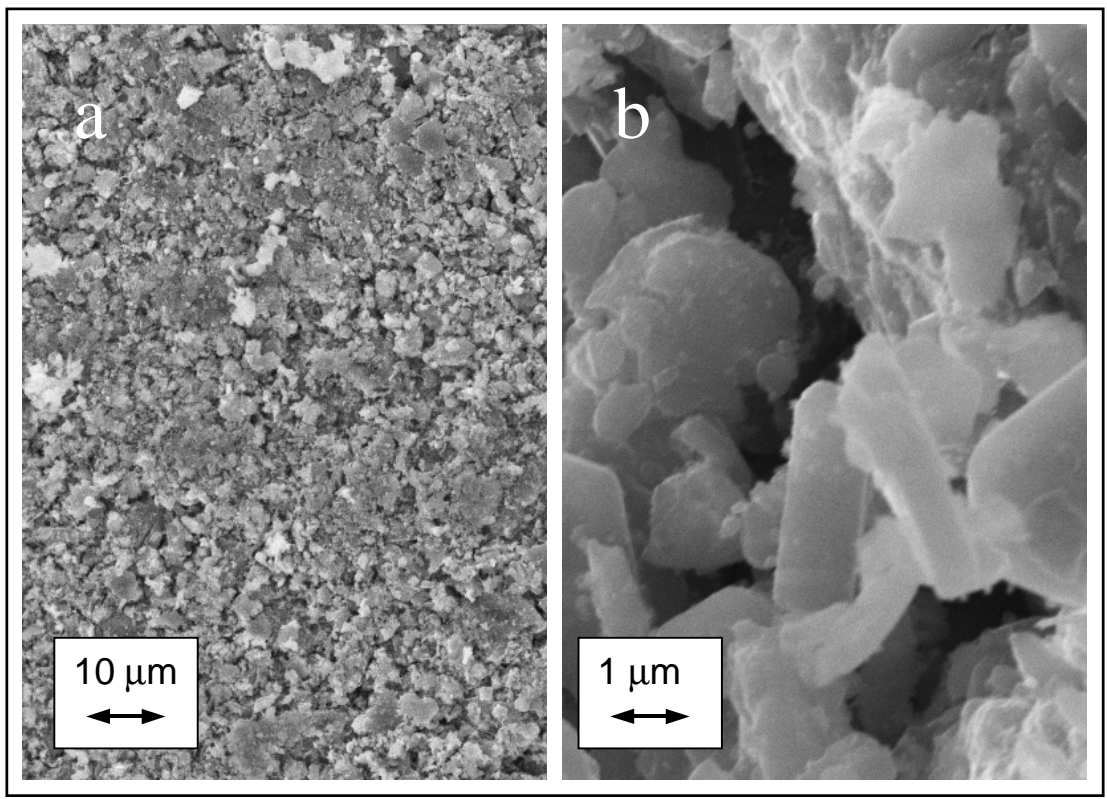

Figure 1: SEM images of the sintered sample.

## 3. The resistivity

The resistivity measurements are presented in fig 2. The critical temperature is $T_c$=38 K, defined at half of the transition, and the amplitude is $\Delta T_c \sim 0.3$ K. The room temperature resistivity has a value of 130 µΩ cm, whereas the resistivity at $T$=40 K is 40 µΩ cm; the residual resistivity ratio value RRR=ρ(273K)/ρ(40K)=3 is nearly the same as those generally found in the literature, even though a much higher RRR value of 20 and resistivity values at room temperature of the order of 10 µΩ cm were reported in ref. [4]. This large difference in the RRR values (which is a more reliable datum, being unaffected by errors in the geometric factors) implies a large difference in the residual resistivity, but this last can be due to both the residual resistivity of the grains and the resistance of the links between the grains. In fact, the presence of impurities (MgO, Mg etc.) at the grain surface can determine a resistance that, to a first approximation, can be thought to be independent of temperature and in series with the resistance of the grains.

Thus, the temperature dependence of the normal state resistivity can be described by the expression:

$$\rho(T) = \rho^{gb} + \rho^{g}(T) = \rho^{gb} + \rho_0^{g} + \rho_{ph}^{g}(T) \tag{1}$$

where $\rho^{gb}$ is the effective resistivity between the grains and $\rho^g(T)$ is the resistivity of the grains which is given by the sum of the residual resistivity, $\rho_0^g$, and of the term due to the scattering with the phonons, $\rho_{ph}^g(T)$. The resistivity curve from 40 to 300 K was best fitted with the function:

$$\rho(T) = \rho_0 + (m-1)\rho'\theta\left(\frac{T}{\theta}\right)^m J_m\left(\frac{T}{\theta}\right) \tag{2}$$

$$J_m\left(\frac{T}{\theta}\right) = \int_0^{\theta/T} \frac{x^m dx}{(e^x - 1)(1 - e^{-x})} \tag{3}$$

where $\theta$ is the Debye temperature, $\rho'$ the temperature coefficient of resistivity for $T >> \theta$, and $m$=3-5. Eq. (2) becomes $\rho(T) = \rho_0 + const \cdot T^m$ for $T < 0.1\,\theta$; therefore, the resistivity being well fitted by a power law $\rho(T) = \rho_0 + const \cdot T^3$ [4] in the range from 40 to 100 K, we have chosen $m$=3, and the best fitting curve obtained with $\rho_0$=39.7 µΩ cm, $\theta$ =1050 K and $\rho'$=0.495 µΩcm/K is reported in fig.2 as a continuous line. The $\theta$ value is in fair agreement with the Debye temperature obtained from heat capacity measurements [15], even though a rather lower value has been also obtained (about 800 K) [16], while the low temperature $T^3$ behaviour, common in transition metals [17], can be related to inter-band scattering processes; the analogy with transition metals suggests that also in $MgB_2$ there is a more mobile band that determines the conduction. $\rho'$, which is proportional to the strength of the electron-phonon coupling, has the same value as in A15 [18], in agreement with a moderate strong coupling found in $MgB_2$. The $\rho_0$ value is quite high, and entails, in the Drude approximation and with a carrier density of $10^{29}$ m$^{-3}$, a carrier mean free path as low as ten Å; therefore, we suggest that $\rho_0$ is mainly due to a large resistance of the grain boundaries.

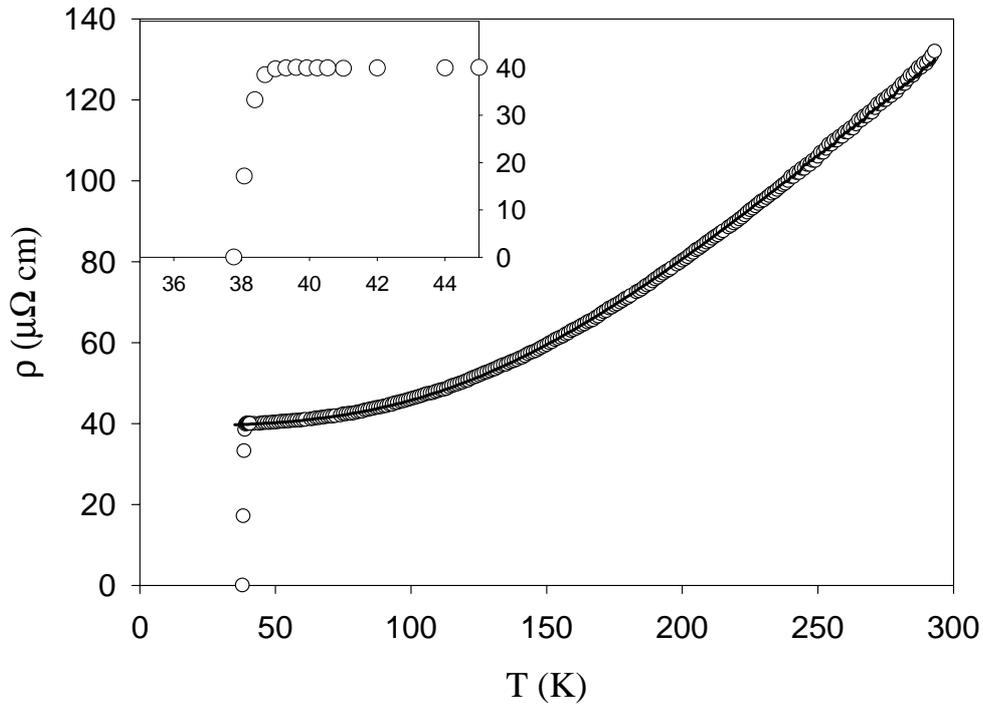

Figure 2: $\rho$ as a function of temperature; the transition region is enlarged in the inset. The best fitting curve, obtained with $m=3$, $\rho_0=39.7\mu\Omega$ cm, $\theta_R =1050$ K, $\rho'=4.95\times10^{-1}$ $\mu\Omega$cm/K, is reported as a continuous line.

## 4. The thermal conductivity

The thermal conductivity measurements are shown in fig.3 from 10 to 280 K. $\kappa$ is about 1 W/mK at 10 K, increases nearly linearly and reaches a broad maximum at about 100 K. Then, it starts to increase again and reaches the value of 9 W/mK at 260 K. In the inset the region from 10 to 50 K is enlarged and $T_c = 37$ K is emphasized, this value being obtained by the simultaneous measure of the Seebeck effect: no anomaly is present around $T_c$. These data are similar to those presented in ref. [8] and [10], where a more pronounced maximum is shown at 100 K and no anomaly at the superconducting transition is present. Such thermal conductivities differ in value and in behaviour from those of simple metals. In fact, the values are rather low considering the high carrier concentration of $MgB_2$, the position of the maximum is at higher temperature than in other metals, and the superconducting anomaly is missing. Actually, the low values of $\kappa$ can be related to the strong electron-phonon coupling, which reduces both phonon and electron mean free path, and the high Debye temperature of this compound could explain the shifting of the peak towards high temperature. The most puzzling aspect is the lack of any kind of anomaly at the superconducting transition. In fact, in the low temperature superconductors, when quasi-particles are the main heat carriers, $\kappa$ decreases below $T_c$ owing to their condensation. Different is the case of the high temperature superconductors, where $\kappa$ shows a peak below $T_c$, explained considering both the increases in the phonon mean free path due to carrier condensation and the increases in the quasi-particles relaxation time in the superconducting state [19]. Before seeking an explanation of the lack of the anomaly it is important that we determine what are the main heat carriers in $MgB_2$ in the normal state.

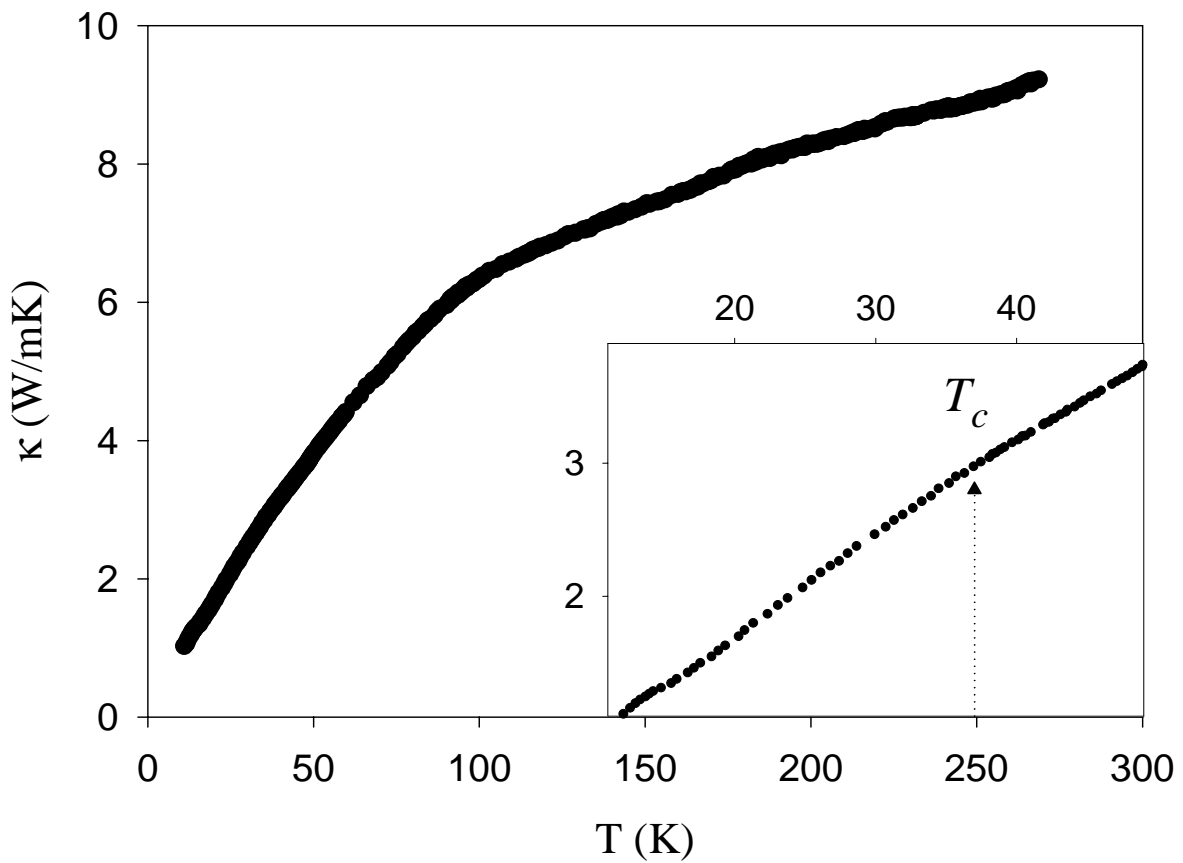

Figure 3: $\kappa$ as a function of $T$; the region from 10 to 50 K is enlarged in the inset, where the critical temperature $T_c = 37\ K$, obtained by the simultaneous measurement of the Seebeck effect, is emphasized.

## 4.1 Thermal conductivity in the normal state

The thermal conductivity in a metal is given by the sum of the electron thermal conductivity, $\kappa_e$, and the phonon thermal conductivity, $\kappa_p$, which we can write as:

$$\kappa_e = \left(W_e^p + W_e^i\right)^{-1} \tag{4}$$

$$\kappa_p = \left(W_p^e\right)^{-1} \tag{5}$$

where, for the Matthiessen's rule, the $\kappa_e$ inverse is the sum of the thermal resistivity for scattering with phonons, $W_e^p$, and for scattering with impurities, $W_e^i$; as for $\kappa_p$, in eq. (5) we assume that, in the temperature range of our interest (10-300 K), it is given only by the inverse of the thermal resistivity for scattering with electrons, $W_p^e$; obviously, this is an overestimate of $\kappa_p$. Following ref. [20] we can write:

$$W_e^i = \frac{\rho_0}{L_0 T} \tag{6}$$

$$W_e^p = \frac{R}{L_0 T}\left(\frac{T}{\theta}\right)^5 J_5\left(\frac{\theta}{T}\right)\left[1 + \frac{3}{\pi^2} n_a^{2/3}\left(\frac{\theta}{T}\right)^2 - \frac{1}{2\pi^2}\frac{J_7(\theta/T)}{J_5(\theta/T)}\right] \tag{7}$$

$$W_p^e = \frac{R}{L_0 T}\left(\frac{T}{\theta}\right) J_5\left(\frac{\theta}{T}\right)\frac{\pi^2 n_a^2}{27[J_4(\theta/T)]^2} \tag{8}$$

where $R = 2\rho'\theta$, $L_0 = 2.44\times10^{-8}$ V$^2$K$^{-2}$ is the Lorenz number, $n_a$ is the number of electrons per atom, and the functions $J_m$ are given by eq. (3). From the resistivity best fit we know the parameters that enter in the Eqs. (6), (7) and (8): they are the Debye temperature $\theta$, the residual resistivity $\rho_0$ related to the residual electron mean free path, and the temperature coefficient $\rho'$.

In fig. 4 $\kappa_e$, $\kappa_e^p = (W_e^p)^{-1}$, $\kappa_e^i = (W_e^i)^{-1}$ and $\kappa_p$, given by eqs. (4)-(8), are plotted with $\rho_0 = 39$ μΩcm, $R = 2A\theta = 1.04\times10^{-5}$ μΩcm, $\theta = 1050$ K, $n_a = 4/3$. First, we can see that $\kappa_p$ is from 3 to 1 order of magnitude less than $\kappa_e$ in the considered temperature range; if we recall that eq. (5) gives an overestimate of $\kappa_p$, we can conclude that phonons give a negligible contribution to the thermal conductivity in MgB$_2$. Second, examining $\kappa_e$ lead us to conclude that it is limited to the scattering with the impurities at low temperature and to the scattering with phonons above 100 K. Now, by comparing the calculated $\kappa_e$ with the measured thermal conductivity, we can see that they look very similar, both in values and in behaviour. The two curves show the same linear increase at the lowest temperatures due to the scattering with impurities, and then the calculated curve shows a peak at about 90 K which is not resolved in the experimental one but which appears clearly in the data of ref. [10].

Finally, we can conclude that the normal state thermal conductivity of $MgB_2$ can be well understood considering that only electrons contribute to it; moreover, the high residual resistivity and the strong electron-phonon coupling may account very well for the low thermal conductivity values compared with those of other metals.

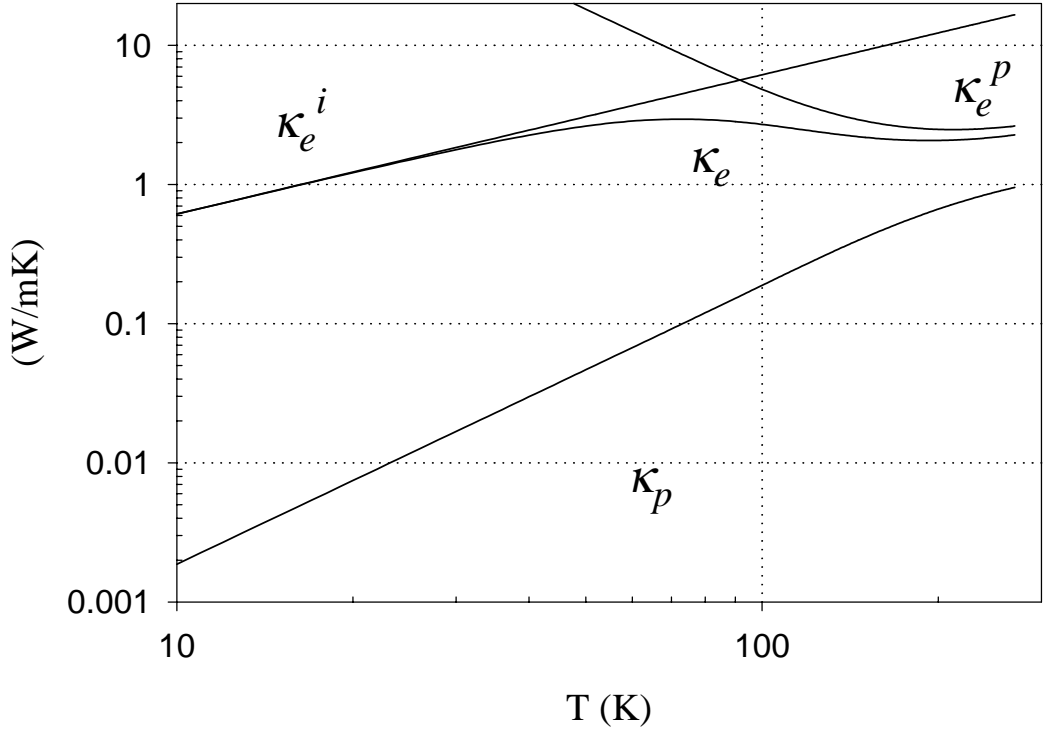

Figure 4: $\kappa_e$, $\kappa_e^p = (W_e^p)^{-1}$, $\kappa_e^i = (W_e^i)^{-1}$ and $\kappa_p$, as functions of temperature; they are given by eqs. (4-8), with $\rho_0$=39 μΩcm, $R = 2A\theta$ =1.04×10$^{-5}$ μΩcm, $\theta$=1050 K, $n_a$=4/3.

### 4.2 Thermal conductivity in the superconducting state

In the following we pursue the hypothesis that electrons are the main heat carriers also in the superconducting state. However, since in the BCS framework the carrier condensation causes the decrease in $\kappa_e$ and the enhancement of $\kappa_p$, this assumption could be false for $t = T/T_c \ll 1$. The electron thermal conductivity in the superconducting state can be written as:

$$\kappa_e^s = \left( \frac{W_e^i}{g_i(t)} + \frac{W_e^p}{g_p(t)} \right)^{-1} \tag{9}$$

The function $g_i(t)$ and $g_p(t)$, calculated in ref. [21] and [22], respectively, are defined as the ratio between the electron thermal conductivity in the superconducting and in the normal state $\kappa_e^s/\kappa_e^n$ in the dirty and in the clean limit, respectively. $g_i(t)$ and $g_p(t)$, which take the condensation of quasi-particles into account, are decreasing function of temperature depending on the amplitude of the energy gap; thus, equation (9) entails a pronounced shoulder below $T_c$ that is not present in our data. Even a close inspection of the data around 37 K did not emphasized any change of slope, as though the superconducting transition did not occur, while the Seebeck effect, measured simultaneously on the same sample, underwent a sharp transition showed in fig. 6.

To explain this puzzle in a purely electron framework, we suppose that the thermal resistance we measure is the series of a normal and a superconducting thermal resistance, and that the first prevails on the latter. The grain boundaries can cause such a normal resistance which does not affect the superconducting transition in resistivity and in Seebeck effect, provided that their thickness is less than the coherence length.

To verify our hypothesis we assume that the superconducting grains are connected to each other through metallic grain boundaries and we schematize such a network as a simple series of thermal resistance:

$$W^{meas} = W_e^g + W_e^{gb} \qquad (10)$$

where $W_e^g$ and $W_e^{gb}$ are the thermal resistance of grains and grain boundaries, respectively. From the resistivity fit we inferred that grain boundaries give rise to a temperature independent resistivity whose value is a fraction $x$ of $\rho_0$: $\rho^{gb} = x\rho_0$. Therefore, the thermal resistance of the grain boundaries will be:

$$W_e^{gb} = \frac{\rho^{gb}}{L_0 T} = \frac{x\rho_0}{L_0 T} \qquad (11)$$

The thermal conductivity of the superconducting grains can be calculated from eqs. (10) and (11):

$$\kappa_e^g = \left\{W^{meas} - W_e^{gb}\right\}^{-1} = \left\{W^{meas} - \frac{x\rho_0}{L_0 T}\right\}^{-1} \qquad (12)$$

$\kappa_e^g$ is plotted in fig. 5 as a function of temperature for $x$=0.6, 0.65, 0.7. As you can see, the shoulder below $T_c$ is now evident and becomes more and more resolved as $x$ increases; also the maximum below 100 K is well resolved. Owing to the roughness of our model, the estimate of the grain boundary resistance is beyond our aims; however, we have demonstrated how efficiently the grain boundaries affect the thermal conductivity, i.e. to the point that they render the superconducting transition invisible.

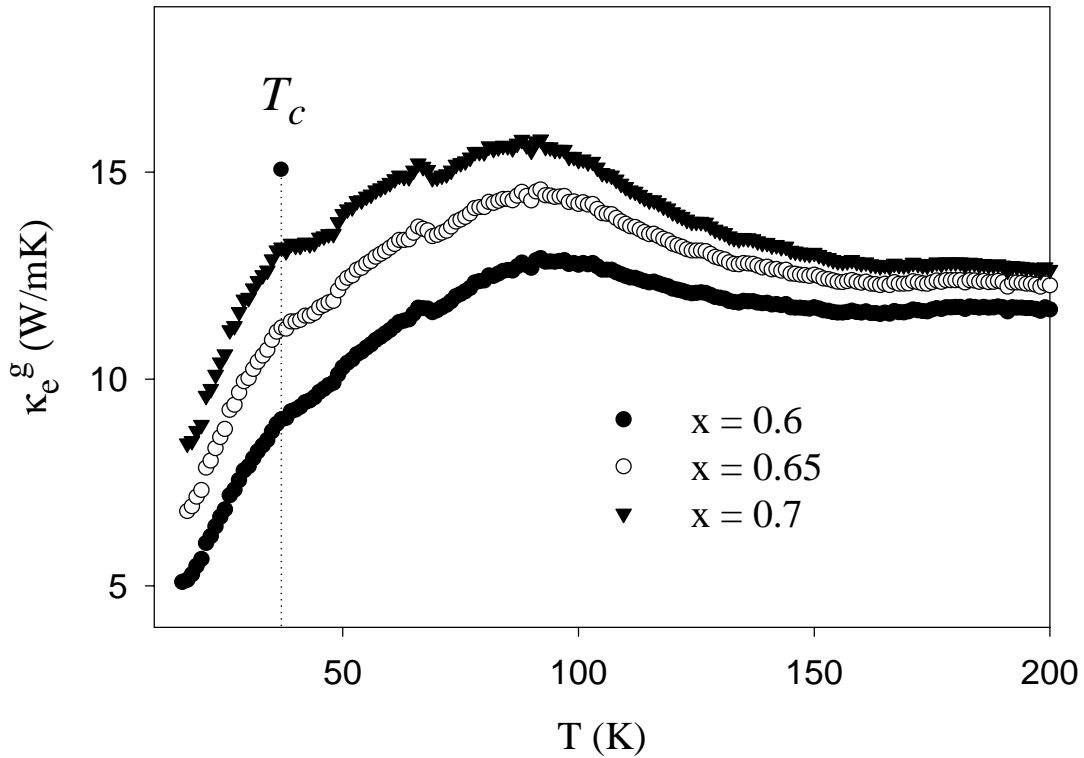

Figure 5: $\kappa_e^g$ as a function of temperature for $x$=0.6, 0.65, 0.7.

## 5. The Seebeck effect

The Seebeck effect measurements are shown in figure 6. The Seebeck effect is zero in the superconducting state, undergoes the transition between 34 and 37 K (see the inset), and shows a small bump just above the transition; above 45 K it starts to increase with a positive curvature that becomes negative above 150 K. The broadening of the transition compared with the resistivity curve, as well as the bump, are probably artifacts due to the a.c. technique which become evident where $S$ varies sharply with temperature [14]. The transition region excepted, the Seebeck effect appears to be in strict agreement with the data presented in the literature [5]-[8] both in value and in behaviour, and this is in contrast with the spreading of resistivity data present in the literature. The small dependence of the Seebeck effect on an extrinsic effect such as disorder or granularity was well proved in cuprate superconductors [13], and this may be taken into account by the Mott formula:

$$S = -\frac{\pi^2}{3}\frac{K_B^2 T}{e}\frac{\sigma'}{\sigma} \qquad (13)$$

where $\sigma$ is the electrical conductivity and $\sigma' = \frac{\partial}{\partial \varepsilon}\sigma(\varepsilon)\Big|_{\varepsilon=\varepsilon_F}$ is the derivative of the electron conductivity with respect to the energy at the Fermi energy $\varepsilon_F$. If the relaxation time, $\tau$, is independent of energy, which is the case for scattering with grain boundaries, $S$ becomes independent of the scattering processes. Equation (13) represents the diffusive contribution of the electrons to the Seebeck effect and gives a linear dependence with temperature. In the isotropic case we have $\sigma'/\sigma=(3/2\varepsilon_F)$ and equation (13) becomes:

$$S = \frac{\pi^2}{2e}\frac{K_B^2 T}{\varepsilon_F} = \frac{C_e}{ne} \qquad (14)$$

where $C_e = \gamma T$ is the electron specific heat per unit volume.

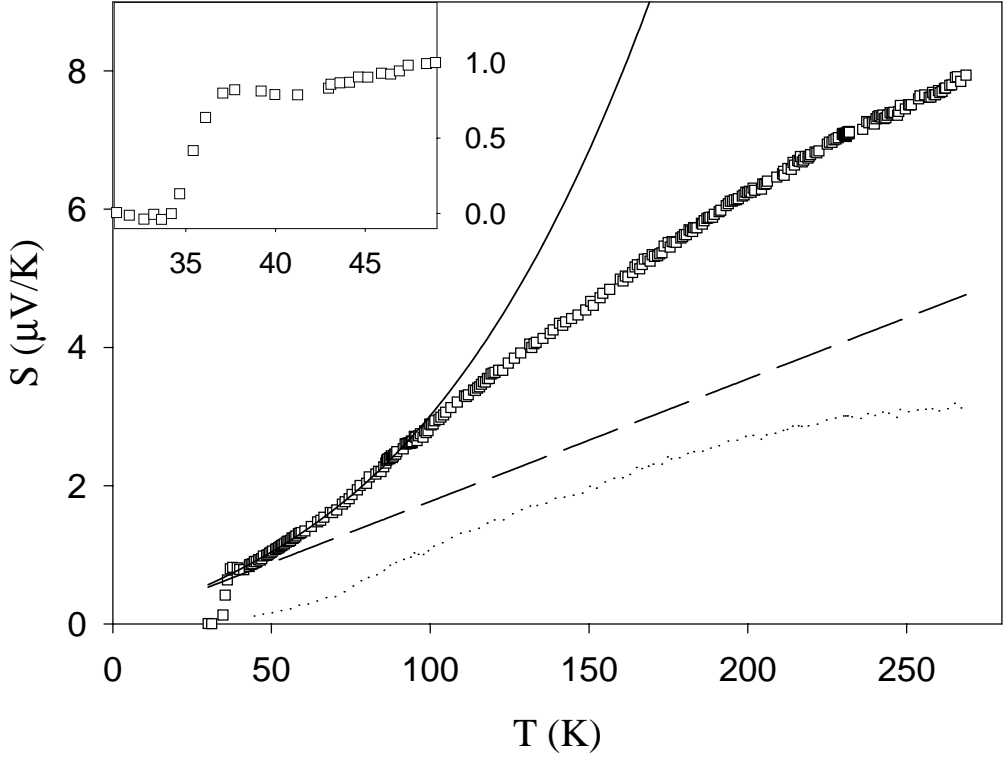

Figure 6: $S$ as a function of temperature; the continuous line is given by eq. (7) with $A=1.76\cdot 10^{-2}$ µV/K$^2$ and $B= 1.26\cdot 10^{-6}$ µV/K$^4$; the dashed line is the diffusive term $S_d=AT$; the dotted line is the experimental phonon drag term defined as $S-S_d$.

In pure metals and for temperatures lower than the Debye temperature the phonon relaxation time for interaction with other phonons and impurities is much longer than the relaxation time for phonon-electron interactions, then phonons do not thermalize themselves and they transfer energy and momentum mainly to the electrons; this mechanism, called phonon drag, gives an extra contribution to the Seebeck effect, known as the phonon drag term, $S_g$, which has to be added to the diffusive term arising from the conduction electrons only:

$$S = S_d + S_g \tag{15}$$

For pure isotropic metals and if the electron-phonon Umklapp processes are neglected, $S_g$ can be written as [23]:

$$S_g = \frac{C_{ph}}{3ne} \tag{16}$$

where $C_{ph}$ is the phonon specific heat per unit volume. Thus, for $T \ll \theta$, the Seebeck coefficient of metals assumes the very simple form:

$$S = S_d + S_g = \frac{C_e}{ne} + \frac{C_{ph}}{3ne} = A \cdot T + B \cdot T^3 \tag{17}$$

where:

$$A = \frac{\gamma}{ne} = \frac{\pi^2 K_b^2}{2e} \frac{1}{\varepsilon_F} \tag{18a}$$

$$B = \frac{\beta_3}{3ne} = \frac{K_b}{e} \frac{1}{n_a} \frac{4\pi^4}{5} \frac{1}{\theta^3} \tag{18b}$$

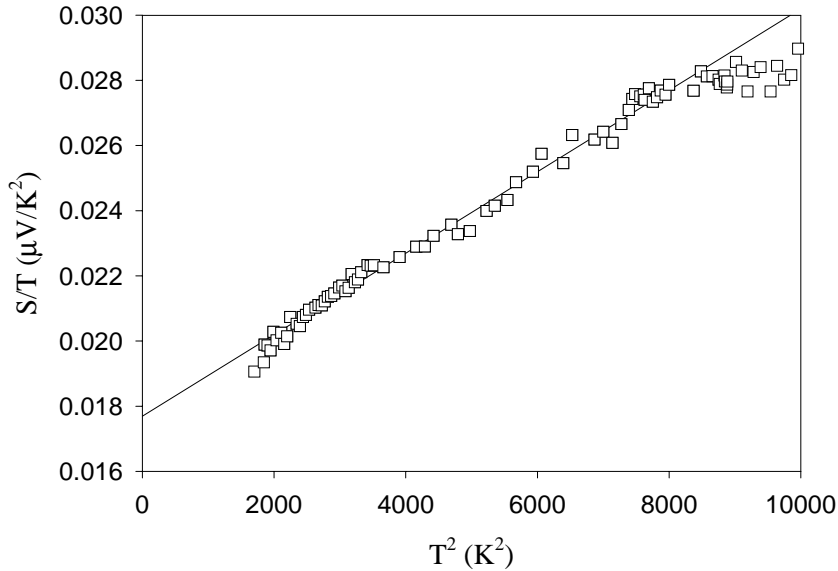

Figure 7: $S/T$ as a function of $T^2$; the continuous line is the best fit performed in the range $2000\ K^2 < T^2 < 8000\ K^2$.

where $\beta_3 = \dfrac{12\pi^4}{5}\dfrac{NK_B}{\theta^3}$ is the coefficient of the low temperature phonon specific heat and N the density of atoms. Equation (17) can be compared with the experimental Seebeck effect of MgB$_2$ in the temperature range T$_C$<T<0.1 $\theta$; this region turns out to be rather extended, considering the large Debye temperature of MgB$_2$.

Fig. 7 shows the ratio S/T as a function of $T^2$ for 40 K <T<100 K. The data show a linear behaviour up to a T$^2$ value of 8000 (T=90 K) and then begin to bend. The best fit performed in the range 2000 K$^2$<T$^2$<8000 K$^2$ is plotted in fig. 6 as a continuous line and the fit parameters are A=1.76·10$^{-2}$ µV/K$^2$ and B= 1.26·10$^{-6}$ µV/K$^4$.

Eq. (17) is plotted as a continuous line in fig. 5, with A and B given by the fit; the diffusive term S$_d$=AT is plotted as a dashed line. As one can see, the experimental curve is well fitted by eq. (17) up to 100 K, then the data change curvature and tend to increase linearly, nearly with the same slope as S$_d$. If we plot the experimental phonon drag term, defined as S-S$_d$ (dotted line), we can see that it tends to saturate above 250 K, which is what we expect for S$_g$; in fact, S$_g$ is proportional to C$_{ph}$ as long as the electron-phonon diffusion is the main scattering mechanism for phonons. As temperature increases, the phonon-impurity and phonon-phonon processes become important, phonons do not transfer momentum only to electrons, and the phonon drag falls. For simple metals as Cu, Ag, Au, Al, this happens at about $\theta$/5 [24] where S$_g$ shows a peak; in the case of MgB$_2$, owing to the strong electron-phonon coupling, the peak has not been reached at $\theta$/4 yet.

Finally, the excellent low temperature fit, obtained in the temperature range where it was expected, and the overall reliable agreement with eq. (15) demonstrate that the diffusive and phonon drag terms contribute nearly to the same extent to the Seebeck effect in the considered temperature range. This was not recognized in ref. [6] and [7], where only a diffusive term that did not vanish at T=0 was considered to fit the data.

Now, we continue to verify the consistency of this model by deriving some microscopic parameters from the coefficients A and B. At first, we refrain from considering the multi-band character of transport properties in MgB$_2$ and we pursue our simple model of isotropic free electrons. The reliability of this model in the light of the band structure will be discussed afterwards.

The coefficients A and B for our sample, together with those extrapolated from data in the literature [5] [6] for pure and Al-doped MgB$_2$, are summarized in table I. The Fermi energy, $\varepsilon_F$, and the Debye temperature, $\theta$, are obtained directly from A and B (eq. (18)) while the electron density n can be obtained from A, once γ=3±1 mJ/moleK$^2$ has been introduced in eq. (18a). For pure MgB$_2$ we find an electron density of 4-8×10$^{28}$ m$^{-3}$, about two times lower than the value obtained from Hall effect measurements [9]; the Debye temperature of about 1400 K, is the only 30% higher than the value found by resistivity measurements. Finally, $\varepsilon_F$ for pure MgB$_2$ is of the order of 2 eV and becomes 1.2 eV for Al doped samples. The decrease in $\varepsilon_F$ with Al doping is interesting: since the Al doping rises the Fermi level, this is a further evidence (with the positive sign) that S is dominated by hole carriers.

**Table I**

|  | A | B | n | θ | $\varepsilon_F$ |
|---|---|---|---|---|---|
|  | μV/K$^2$ | μV/K$^4$ | $10^{28}$ m$^{-3}$ | K | eV |
| MgB$_2$ | 1.76×10$^{-2}$ | 1.26×10$^{-6}$ | 6.±2 | 1430 | 2.08 |
| [a]MgB$_2$ | 1.70×10$^{-2}$ | 1.30×10$^{-6}$ | 6.±2 | 1430 | 2.17 |
| [b]MgB$_2$ | 2.00×10$^{-2}$ | 1.30×10$^{-6}$ | 5.±2 | 1430 | 1.82 |
| [b]Mg$_{0.95}$Al$_{0.05}$B$_2$ | 2.88×10$^{-2}$ | 1.41×10$^{-6}$ |  | 1390 | 1.27 |
| [b]Mg$_{0.9}$Al$_{0.1}$B$_2$ | 2.99×10$^{-2}$ | 1.49×10$^{-6}$ |  | 1360 | 1.22 |

[a]Ref. [5], [b]Ref. [6]

Now, let us consider some details of the band structure of MgB$_2$. The bands that contribute to the conduction are of two types [3][25]: two σ bands, deriving from the $p_{x,y}$ states of B, and two π bands, deriving from the $p_z$ states. The two sets of bands have very different character, the σ band being of hole-type and nearly 2D (there is very little dispersion in the Γ – A direction), and the π band being mainly of electron-type and 3D. In the discussion of the pairing mechanism great relevance has been given to the σ bands [2][3][26]. The positive sign of the Hall coefficient and of the TEP as well as the increase in the latter with Al doping confirm the importance of these bands in the transport properties. Thus, assuming that only these bands contribute to the TEP, we must calculate it taking the 2D nature of σ bands into account, which entails that S is a tensor. In the hypothesis of a cylindrical Fermi a surface running along Γ-A and of a scattering relaxation time independent of energy, the Mott formula (3) gives in this case:

$$S_{\sigma xx} = S_{\sigma yy} = \frac{\pi^2}{3e} \frac{K_B^2 T}{\varepsilon_{F\sigma}} \qquad (19a)$$

$$S_{\sigma zz} = 0 \tag{19b}$$

To compare eqs. (19) with the experimental data, taken on polycrystalline samples, we ought to make an average with respect to the three directions:

$$\overline{S\sigma} = \frac{2}{9}\frac{\pi^2}{e}\frac{K_B^2 T}{\varepsilon_{F\sigma}} \tag{19c}$$

and similarly for the phonon drag term.

$$\overline{S_g} = \frac{K_b}{e}\frac{1}{n_a}\frac{8\pi^4}{15}\left(\frac{T}{\theta}\right)^3 \tag{20}$$

The expression (19c) differs from eq. (13) only by the numerical factor 4/9, while eq. (20) differs from eq. (16) by a factor 2/3. The Fermi energy of the σ band, $\varepsilon_{F\sigma}$, calculated by comparing $\overline{S_\sigma}/T$ with the coefficients $A$ and the Debye temperature calculated by comparing $\overline{S_g}/T$ with $B$, are reported in table II. In this way we find a value for pure $MgB_2$ that is in fair agreement with the difference between the Fermi energy and the top of the s bands $\varepsilon_{top} =$ 0.9 eV calculated by Suzuki [26]; the same paper calculated the reduction of $\varepsilon_{top}$ with Al doping, which turned out to amount to about 15% for $Mg_{0.9}Al_{0.1}B_2$, two times lower than what we obtain.

The Debye temperature values in tab. II are smaller than those in tab. I and, therefore, in better agreement with the value found from specific heat and resistivity; the decreasing of the $\theta$ value with the Al doping can be due to a softening of the phonon frequencies. Also these results confirm that the hypothesis of 2D conduction is suitable for $MgB_2$.

**Table 2**

|  | $\varepsilon_{F\sigma}$ | $\theta$ |
|---|---|---|
|  | eV | K |
| $MgB_2$ | 0.92 | 1260 |
| [a]$MgB_2$ | 0.96 | 1260 |
| [b]$MgB_2$ | 0.81 | 1250 |
| [b]$Mg_{0.95}Al_{0.05}B_2$ | 0.57 | 1210 |
| [b]$Mg_{0.9}Al_{0.1}B_2$ | 0.54 | 1190 |

[a]Ref. [5], [b]Ref. [6]

## 6. Conclusion

In summary, we analyzed in detail the resistivity, the thermal conductivity and the thermoelectric power in the normal state and we showed that these transport properties are affected in a different way by granularity. We have then showed that all these properties can described within an independent electron framework, by taking into account the high phonon frequencies and the strong electron-phonon coupling. In particular the TEP is the sum of a diffusive and a phonon drag terms which contribute nearly equally to it. The phonon drag term comes out larger in $MgB_2$ than in other metals and this can be due to the importance of electron-phonon interaction in this compound. The diffusive term is positive and increases with Al doping. Considering a simple model in which the conduction is dominated by the 2D $\sigma$ bands, from the comparison of the experimental values with the theoretical ones, we obtain reliable values of the Fermi energy and of the Debye temperature with the increasing of the Al doping. Further investigation will be necessary to verify this result; in particular transport properties on samples with higher level of Al doping will be a useful tool to better investigate the role of the $\sigma$ bands, which have great importance in the pairing mechanism.

**References**

[1] J.Nagamatsu, N.Nakagawa, T.Muranaka,Y.Zenitani and J.Akimitsu, Nature **410**, (2001) 63.
[2] J.E. Hirsch, cond-mat /0102115; J.E.Hirsch and F.Marsiglio cond-mat/0102479.
[3] J.Kortus, I.I.Mazin, K.D.Belashenko, V.P.Antropov, L.L.Boyer, cond-mat/0101446.
[4] P.C.Canfield, D.K.Finnemore, S.L.Bud'ko, J.E.Ostenson, G.Lapertot, C.E.Cunningham, C.Petrovic, cond-mat/0102289.
[5] B. Lorenz, R. L. Meng, C. W. Chu, cond-mat/0102264.
[6] B. Lorenz, R. L. Meng, Y. Y. Xue, C. W. Chu, cond-mat/0104041.
[7] E. S. Choi, W. Kang, J.Y.Kim, Min-Seok Park, C.U.Jung, Heon-Jung Kim, Sung-Ik Li, cond-mat/0104454
[8] M.Schneider, D.Lipp, A.Gladun, P.Zahn, A.Handstein, G.Fuchs, S.L.Drechsler, M.Richter, K.-H.Mueller, H.Rosner, cond-mat/0105429.
[9] V.N.Kang, C.U.Jung, Kijon H.P.Kim, Min-Seok Park, S.Y.Lee, Hyeong-Jin Kim, Eun-Mi Choi, Kyung Hee Kim, Mun-Seog Kim, Sung-Ik Lee, cond-mat/0102313; V.N.Kang, Hyeong-Jin Kim, Eun-Mi Choi, Kijon H.P.Kim, Sung-Ik Lee, cond-mat/0105024; R.Jin, M.Paranthaman, H.Y.Zhai, H.M.Christen and D.Mandrus, cond-mat/0104411.

[10] E.Bauer, Ch. Paul, St. Berger, S.Majumdar, H.Michor, M.Giovannini, A.Saccone, A.Bianconi, cond-mat/014203.

[11] M.Kambara, N.Hari Babu, E.S.Sadki, J.R.Cooper, H.Minami, D.A.Cardwell, A.M.Campbell, I.H.Inoue, Supercond. Sci. Technol. 14 (2001) L5.

[12] M. R. Cimberle, M. Novak, P. Manfrinetti, A. Palenzona, cond-mat/0105212.

[13] J.L.Tallon, J.R.Cooper, P.S.I.P.N. de Silva, G.V.M.Williams and J.W.Loram, Phys.Rev.Lett. **75** (1995) 4114.

[14] M.Putti, A.Canesi, M.R.Cimberle, C.Foglia, A.S.Siri, Phys. Rev. B **58** (1998) 12344.

[15] F.Bouquet, R.A.Fisher, N.E.Phillips, D.G.Hinks, J.orgensen, cond-mat/0104206

[16] R.K.Kremer, B.J.Gibson, K.Ahn, cond-mat/0102432

[17] G.T. Meaden, *Electrical resistance of metals,* p. 89, Heywood Books, London, (1966).

[18] F.Nava, O.Bisi, K.N.Tu, Phys. Rev. B **34** (1986) 6143.

[19] S.Castellazzi, M.R.Cimberle, C.Ferdeghini, E.Giannini, G.Grasso, D.Marrè, M.Putti, A.S.Siri, Physica C **273** (1997) 314.

[20] R.Berman, *Thermal conduction in solids,* p. 133, Clarendon Press, Oxford, (1976).

[21] J.Bardeen, G.Rickayzen, L.Tewordt, Phys. Rev. **133** (1959) 982.

[22] B.T.Geilikman, M.I.Dushenat, V.R.Chechetkin, Sov. Phys. JEPT 46 (1977) 1213.

[23] J.M.Ziman, *Electrons and phonons*, Clarendon Press, Oxford (1960); R.D.Barnard, *Thermoelectricity in metals and alloys*, Taylor & Francis Ltd, London, 136 (1972).

[24] Landolt-Börnstein, Volume 15 b *Metals: electronic transport phenomena* Editors: K.-H. Hellwege and J.L.Olsen, Springer -Verlag (1985).

[25] G.Satta, G.Profeta, F.Bernardini, A.Continenza, S.Massidda, cond-mat/0102358.

[26] J.M.An, E.W.Pickett,Phys. Rev. Lett. **86** (2001) 4366..

[27] S.Suzuki, S.Higai, K.Nakao, cond-mat/0102484.